# Development of embedded target detection system based on FPGA and YOLOv3-Tiny

Zihan Jiang [a, 1],Fanghao Liu [a, 1], Huawei Wang [a], Mamataziz Mattohti [a], Xiangquan Chen [a]

Jingfu Guo [a, b, *], Xiaotian Wu [a, b, *], Yongjun Dong [a, b, *]

[a] School of Physics, Northeast Normal University, Changchun 130024, China.

[b] Key Laboratory of Advanced Energy Development and Application Innovation under Jilin Province, Changchun 130024, China

**Abstract:** Computational complexity and storage requirements are crucial factors influencing the performance and efficiency of convolutional neural networks (CNNs) in resource-constrained environments. This paper presents a high-performance embedded target detection system based on FPGA and YOLOv3-Tiny, specifically designed for embedded artificial intelligence applications. By integrating lightweight CNN optimization techniques with hardware accelerator design, significant improvements are made in both computational efficiency and resource utilization. Key optimizations, including low-bit quantization, batch normalization fusion, and table lookup mapping, reduce model parameters and computational complexity. Additionally, an FPGA hardware accelerator with a pipelined architecture is developed to enhance the efficiency of convolution operations while minimizing off-chip data transmission through modular design and on-chip cache optimization. On the ZYNQ-XC7Z035 platform, the system achieves an inference latency of 0.211 seconds, outperforming comparable designs by 75.58% in speed. The system achieves an power efficiency of 10.11 GOPS/W, surpassing comparable designs by at least 29.45%. Furthermore, hardware resource utilization is reduced by up to 51.94% compared to similar systems. This study offers innovative design methodologies and practical application examples for the efficient deployment of deep learning models on embedded platforms.

**Key words:** FPGA; YOLOv3-Tiny; Object detection; Hardware acceleration; Embedded system

## 1 INTRODUCTION

In recent years, convolutional neural networks (CNNs) have achieved remarkable success in image classification, object detection, natural language processing, and speech recognition. Notably, CNN-based object detection technologies have become a prominent research focus, especially in applications such as autonomous driving, intelligent surveillance, drone navigation, and robotic vision[1]. However, deploying such technologies on embedded platforms remains challenging. The high computational

---

* Corresponding author.

E-mail addresses: jiangzh@nenu.edu.cn(Zihan Jiang), 752118410@qq.com(Fanghao Liu), wanghuawei@nenu.edu.cn(Huawei Wang),maimtaz546@nenu.edu.cn(Mamataziz Mattohti), chenxq084@nenu.edu.cn(Xiangquan Chen), guojf217@nenu.edu.cn(Jingfu Guo),wuxiaotian@nenu.edu.cn(Xiaotian Wu), dongyj512@nenu.edu.cn(Yongjun Dong).

[1] These authors contributed equally to this work.

complexity and storage demands of CNNs impose stringent requirements on processor performance, memory bandwidth, and power consumption. These requirements conflict with the inherent resource and power constraints of embedded systems, necessitating the design of efficient, low-power embedded object detection systems based on CNNs.

To address the deployment challenges of object detection algorithms on embedded platforms, researchers globally have explored two primary approaches: algorithm optimization and hardware acceleration. Field-programmable gate arrays (FPGAs), with their high parallelism, flexibility, and low power consumption, offer a promising solution for embedded object detection systems[2].In algorithm optimization, key strategies include lightweight network architecture design, model pruning, quantization, and efficient feature extraction. For instance, [3] proposed a lightweight face detection network based on the YOLO algorithm, which significantly enhanced detection speed and real-time performance while maintaining accuracy. [4] introduced an improved lightweight Mobilenetv3-tiny network that balances detection speed and precision while maintaining its suitability for deployment. For hardware acceleration, [5] developed an FPGA-based object detection system implementing YOLOv2, which enhances real-time performance and hardware efficiency through optimized floating-point matrix multiplication units and dual-buffer data processing circuits. [6] proposed a CPU-GPU-CPU multi-core cooperative strategy to optimize YOLOv5s performance on embedded platforms. [7] achieved a threefold improvement in single-thread inference efficiency on RISC-V architectures via memory access optimization and vectorization. Compared to GPUs and ASICs, FPGAs are widely recognized for their low power consumption and flexibility, yet further integration of lightweight network optimization with hardware acceleration remains underexplored.

To meet embedded object detection demands, this paper introduces an FPGA-based solution with lightweight CNNs. We first reduce computation and resource usage by lightweight YOLOv3-Tiny redesign, low-bit quantization, and optimized data transfer. A pipelined CNN accelerator is then designed, using modular architecture and memory optimization to boost inference speed and energy efficiency. Experiments validate the system's performance, resource efficiency, and power economy, providing a practical framework for embedded detection.

## 2 OVERALL DESIGN SCHEME

This paper designs an embedded object detection system based on Xilinx ZYNQ-XC7Z035. The overall architecture of the system can be divided into two parts: the image acquisition and display system and the convolutional neural network accelerator. The collaborative design of software and hardware enhances both inference performance and energy efficiency.

### 2.1 System Architecture

The system design framework is shown in Figure 1, mainly including the camera initialization configuration, image acquisition, convolutional neural network hardware accelerator, HDMI display, and other modules. The system design fully leverages the collaborative computing capabilities of the PS (Processing System) and PL (Programmable Logic) components of the ZYNQ-XC7Z035. The PS component is responsible for initializing the CMOS image sensor and managing task scheduling, while the PL component handles image acquisition, processing, and accelerated computing. The acquired image data follows two paths: one path transmits the data to the DDR3 mounted on the PS component via AXI VDMA for buffering, then outputs it to the display through the HDMI interface for real-time display; the other path scales the data to fit the input size of the YOLOv3-Tiny network, buffers it in DDR3, and then sends it to the convolutional neural network hardware accelerator via AXI DMA for forward computation. The target prediction results, including detection box coordinates and other information, are integrated with the original image and displayed through the HDMI interface. The system fully combines the features of the Zynq SoC architecture, utilizing the high parallelism of the PL component's logic resources to handle complex computational tasks, while leveraging the high flexibility of the PS ARM processor for task scheduling. Through the reasonable division of tasks, the system operates efficiently and stably.

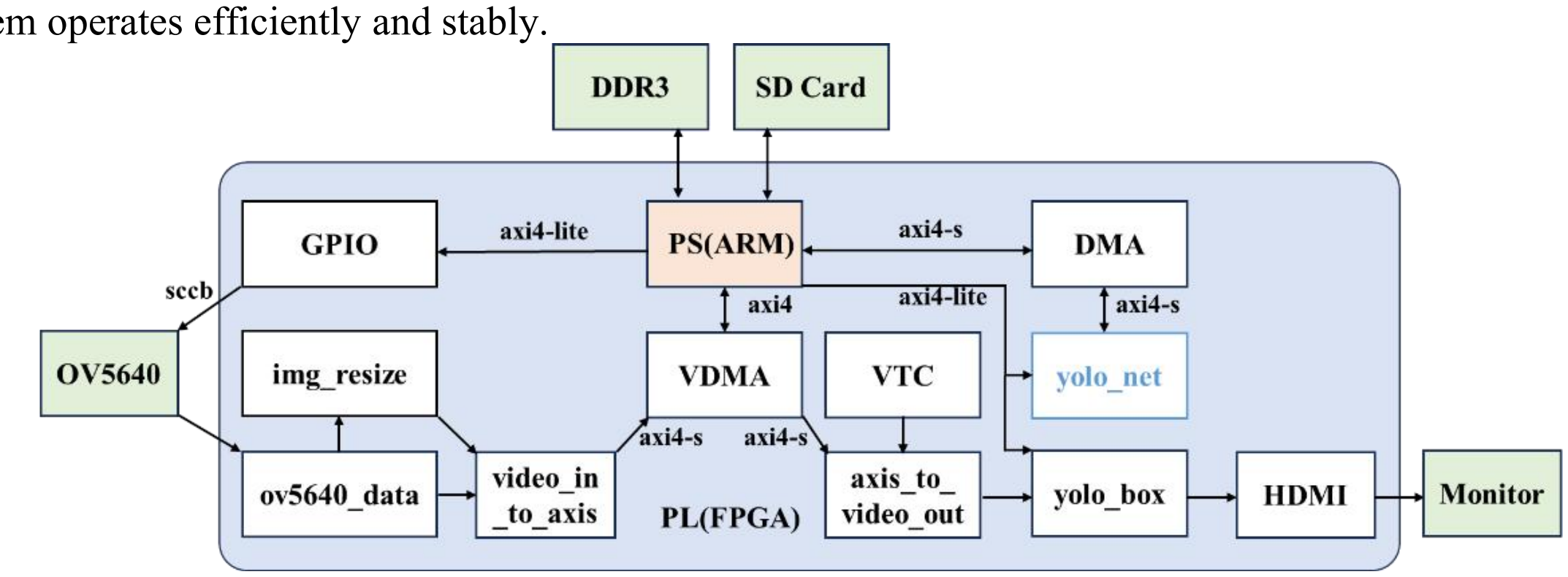

Figure 1: Design Framework of the Target Detection System.

### 2.2 Functional Requirements

The embedded target detection system designed in this paper must meet three major requirements: high efficiency, low power consumption, and resource optimization, to adapt to the constraints of the embedded environment. Firstly, the system must implement the inference operation of the YOLOv3-Tiny network on the ZYNQ-XC7Z035 platform, with a significant advantage in inference speed compared to other design methods, ensuring real-time performance in object detection. Secondly, the system must strictly control power consumption, optimize circuit design and data transmission processes, and maximize performance output within limited power consumption. Finally, considering the limited resources of the FPGA, the hardware architecture is optimized through pipeline parallel computing and memory access strategies, aiming to maximize on-chip resource utilization and minimize off-chip memory usage. The overall design goal is to achieve high efficiency and low power consumption for embedded object detection while ensuring accuracy.

## 3 OPTIMIZATION OF YOLOV3-TINY

The YOLO (You Only Look Once) algorithm is an efficient one-stage object detection framework. Compared to two-stage algorithms, YOLO achieves faster detection speeds, making it suitable for real-time applications such as autonomous driving, intelligent surveillance, and drone navigation. YOLOv3-Tiny, a lightweight variant of YOLOv3, is designed for resource-constrained environments. By simplifying the network structure, it reduces computational demands and improves inference speed, enabling deployment on embedded devices with low power consumption and high real-time performance. The architecture of YOLOv3-Tiny is illustrated in Figure 2.

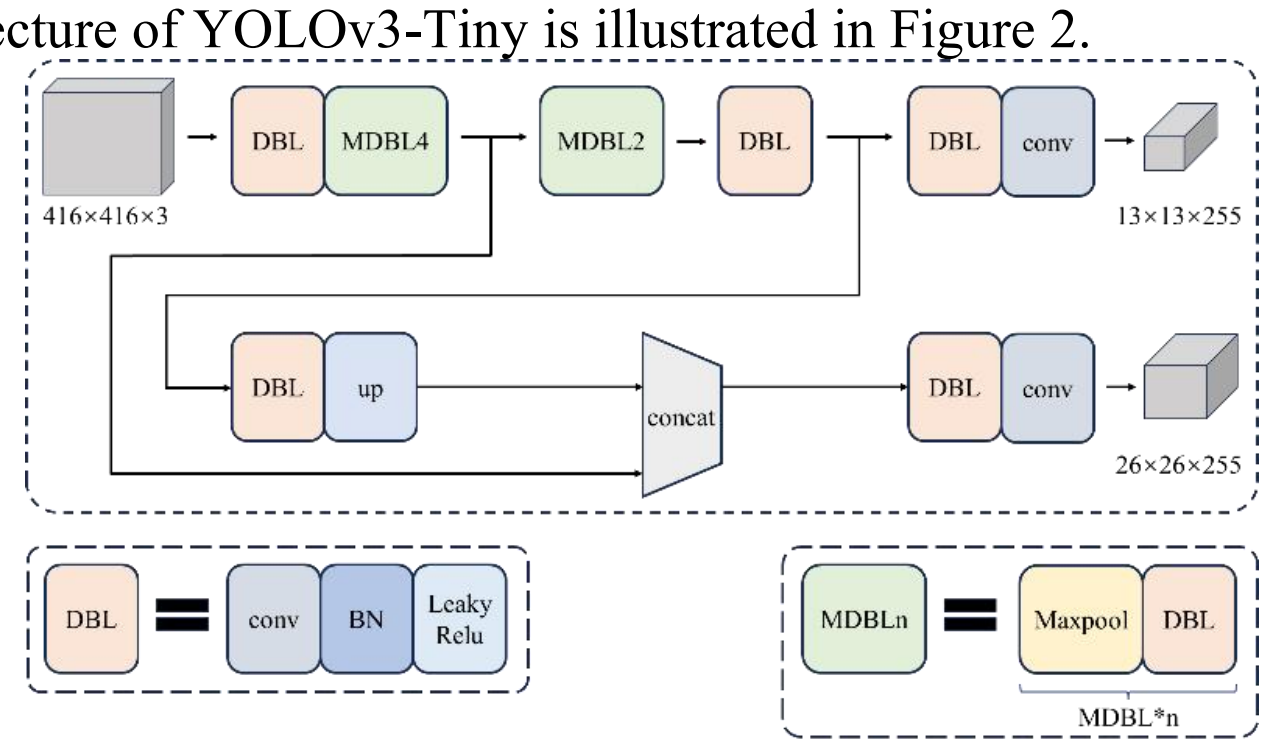

Figure 2: YOLOv3-Tiny network architecture.

### 3.1 Quantization of Network Parameters

Network parameter quantization is an effective model compression technique widely used in hardware acceleration of deep neural networks. By quantizing the floating-point weights and bias parameters of the network model to lower-width fixed-point numbers, storage requirements and computational complexity are reduced, making it suitable for convolutional neural network deployment on embedded platforms such as FPGA. Traditional convolutional networks typically use 32-bit floating-point weights, occupying significant storage space. Since FPGA's on-chip storage is limited and lacks floating-point computation hardware, quantization is required to convert floating-point numbers to 8-bit fixed-point values. This paper uses a linear mapping quantization method[8], which can be expressed as:

$$r = S(q - Z) \tag{1}$$

In the quantization equation, $r$ represents the original floating-point number, and $q$ denotes the quantized fixed-point number. Here, $S$ (Scale) is the scaling factor, and $Z$ (Zero-Point) is the zero-point offset. The scaling factor $S$ and zero-point offset are computed as follows.

$$S = \frac{r_{\max} - r_{\min}}{q_{\max} - q_{\min}} \tag{1}$$

$$q = round(\frac{r}{S} + Z) \tag{2}$$

The scaling factor $S$ scales floating-point values to the fixed-point range, while the zero-point offset $Z$ compensates for the offset between the floating-point and integer ranges. The target quantized fixed-point value $q$ is assigned an 8-bit precision. For unsigned quantization, $q$ ranges within $[0,255]$; for signed quantization, its range is $[-128,127]$.

In YOLOv3-Tiny, there are 13 convolution layers, and multiply-accumulate operations are the primary computational load. Hardware deployment on the ZYNQ-XC7Z035 requires quantization for these operations. When multiplying two floating-point numbers, the quantization process is as follows:

$$S_3(q_3 - Z_3) = S_1(q_1 - Z_1)S_2(q_2 - Z_2) + bias \tag{3}$$

The input image data and weight parameters are multiplied and then added to the bias parameter, which is an important step in the calculation process. From equations (1) and (4), the floating-point values $r_1$ and $r_2$ represent the input image data and weight parameters, respectively, while the bias parameter is added. The result is the floating-point value $r_3$, which can be simplified as:

$$q_3 = \frac{S_1 S_2}{S_3}[(q_1 - Z_1)(q_2 - Z_2) + \frac{bias}{S_1 S_2}] + Z_3 \tag{4}$$

Equation (5) provides the fixed-point computation result of the input-weight product combined with the bias term. On the right-hand side, $q_1$, $q_2$, $Z_1$, $Z_2$, and $Z_3$ are all fixed-point values. Since the scaling factor $S$ is a small value close to zero, the term $bias/(S_1 S_2)$ is also quantized as a fixed-point number. Here, the scaling factor for the bias term is defined as:

$$S_{bias} = S_1 S_2 \tag{5}$$

Due to $\frac{S_1 S_2}{S_3}$ being a very small floating-point number, the following holds:

$$\frac{S_1 S_2}{S_3} = 2^{-n} M_0 \tag{6}$$

where $n$ is an integer, and

$$M_1 = 2^{15} M_0 \tag{7}$$

Thus, from equations (7) and (8), we have:

$$\frac{S_1 S_2}{S_3} = 2^{-(n+15)} M_1 \tag{8}$$

From equations (5) and (9), it can be concluded that the fixed-point result of multiplying input image data and convolution weight parameters, followed by adding the bias, transforms the product of two floating-point numbers into a form where the sum is calculated with a third floating-point number, then quantized into fixed-point values. Additionally, the decimal calculation is optimized into a power-of-2 operation, allowing the FPGA hardware to perform complex computations through simple shift operations, significantly reducing computational complexity. The number of shifts is determined by the quantization scaling factor.

In tests with multi-resolution input images, the inference speed and average precision (AP0.5) comparison between the floating-point and quantized models are shown in Figure 3. As the input resolution increases, the number of pixels the model needs to process also increases, resulting in higher inference latency. In contrast, the quantized model significantly outperforms the floating-point model in terms of latency, with only a slight decrease in accuracy. Experimental results show that the quantized model reduces parameter storage by 75%, shortens forward computation time by 29.05%, and improves efficiency significantly, with only a 10.61% loss in average precision. Overall, the quantized model offers an excellent trade-off between latency and accuracy, closely matching the floating-point model while providing substantial efficiency advantages. As observed from the trend line in the figure, the quantized model offers a good balance between average precision and latency, making it suitable for real-time detection tasks. This optimization provides an efficient and feasible solution for deep learning deployment on embedded devices.

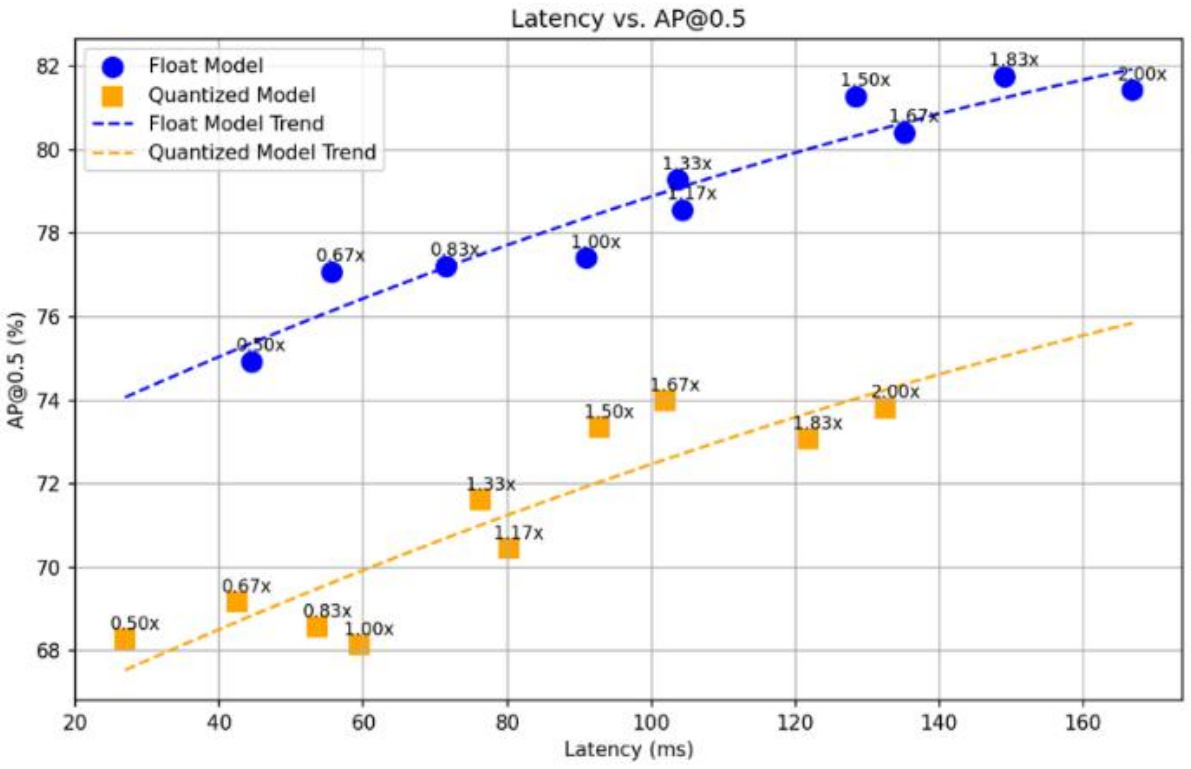

Figure 3: Comparison of Accuracy and Latency Before and After Quantization.

### 3.2 Batch Normalization Layer Fusion

Batch Normalization (BN) is a commonly used technique in deep learning. By normalizing the activation values of network layers, it enhances the stability and convergence speed of training while alleviating the problems of vanishing or exploding gradients. It has become an essential component of modern network models. In convolutional neural networks, BN layers are typically recommended to be inserted before each activation layer. The core computation includes calculating the mean and variance, followed by normalization, scaling, and shifting. The specific process is as follows:

$$\mu_{\mathcal{B}} = \frac{1}{m}\sum_{i=1}^{m} x_i \tag{9}$$

$$\sigma_{\mathcal{B}}^2 = \frac{1}{m}\sum_{i=1}^{m}(x_i - \mu_{\mathcal{B}})^2 \tag{10}$$

$$\hat{x}_i = \frac{x_i - \mu_{\mathcal{B}}}{\sqrt{\sigma_{\mathcal{B}}^2 + \epsilon}} \tag{11}$$

$$y_i = \gamma \hat{x}_i + \beta \tag{12}$$

From equations (10) to (13), the output of the batch normalization can be expressed as:

$$y_i = \gamma \frac{x_i - \mu_B}{\sqrt{{\mu_B}^2 + \epsilon}} + \beta \tag{13}$$

Although the BN layer improves training stability and convergence speed, its standardization operation increases computational overhead, making its implementation on FPGA hardware more challenging. Therefore, BN computation and the weight and bias parameters in convolution layers can be fused to simplify the computation during inference. The convolution computation formula is as follows:

$$Y_k^j = \sum_{i=1}^{C} x_i^j \times w_{k,i}^j + b \tag{14}$$

Here, $x_i^j$ represents the feature map passed through the $i$ -th convolution, $w_{k,i}^j$ represents the $k$ -th filter of the $j$ -th convolution layer, and $Y_k^j$ represents the output data of the $k$ -th convolution layer. Combining equations (10) to (15), the fused convolution and BN computation can be written as:

$$Y_k^j = \sum_{i=1}^{C} x_i^j \times \frac{\gamma \times w_{k,i}^j}{\sqrt{\delta_Y^2 + \varepsilon}} + \frac{\gamma \times (b - \mu_Y)}{\sqrt{\delta_Y^2 + \varepsilon}} + \beta \tag{15}$$

Through the fusion of BN and convolution layers, the new weight and bias parameters are computed as follows:

$$w' = \frac{\gamma \times w_{k,i}^j}{\sqrt{\delta_Y^2 + \varepsilon}} \tag{16}$$

$$b' = \frac{\gamma \times (b - \mu_Y)}{\sqrt{\delta_Y^2 + \varepsilon}} + \beta \tag{17}$$

With the above fusion method, the convolution layer computation can be completed without the need for additional BN layer operations. This approach not only maintains the computational accuracy of the model but also significantly reduces the number of computations and memory accesses. Furthermore, as intermediate results no longer need to be cached, a large amount of storage resources is saved, further optimizing the FPGA hardware deployment performance. Experimental results show that the fused weight and bias parameters can maintain the performance of the original model, while significantly reducing hardware resource usage, providing an efficient solution for embedded object detection systems.

### 3.3 Table Lookup Mapping Acceleration

In the FPGA deployment of the YOLOv3-Tiny convolutional neural network, the implementation of functions such as exponentials and piecewise functions is challenging due to the limitations of FPGA's lower-level resources. These functions often require significant amounts of sequential computation and complex mathematical operations, which consume a large amount of hardware resources such as multipliers and adders, while also increasing computational latency. Therefore, table lookup is an effective optimization technique. By precomputing the function values within a possible input range and storing them in the on-chip memory, it avoids the extra cost of real-time computation during inference by directly looking up the results from the table.

The activation function in the YOLOv3-Tiny network is used after the fusion of the convolution layer and BN layer, and it employs the *LeakyRelu* activation function, expressed as:

$$LeakyRelu(x) = \begin{cases} x & x > 0 \\ ax & x \leq 0 \end{cases} \tag{18}$$

where the slope coefficient $\alpha$ is manually assigned based on prior knowledge (typically $\alpha$ =0.01). During inference, the quantized fixed-point inputs must first be dequantized to floating-point values. After comparing it with the zero point, the activation value is calculated by multiplying it with the slope $\alpha$, and the final floating-point product is quantized into a fixed-point number.

Since the FPGA part of the ZYNQ-XC7Z035 is inefficient at handling floating-point operations, and the input data range of the activation layer is relatively small, the 256 possible input values corresponding to 8-bit fixed-point numbers can be pre-set. This allows the optimization of the computation process through table lookup mapping. The table lookup method precomputes the activation output values for these 256 input values and stores them in the on-chip memory of the FPGA. During inference, the FPGA directly queries the activation values based on the input data address, replacing the complex floating-point calculations. This significantly improves computation efficiency and reduces hardware resource usage.

Through the optimization of table lookup mapping, the implementation of the activation function shifts from relying on real-time computation to simple memory access operations. This reduces both computational latency and hardware resource demands, providing crucial support for the efficient deployment of YOLOv3-Tiny on FPGA.

## 4 FPGA HARDWARE ACCELERATOR DESIGN

### 4.1 Hardware Architecture Design

The YOLOv3-Tiny network model consists of convolution layers, activation layers, max-pooling layers, and upsampling layers, with multiple calls to different network layers, resulting in high computational complexity. Given the limited hardware resources of the ZYNQ-XC7Z035 platform, this paper leverages the parallel computation advantages of the PL (Programmable Logic) side and the flexible processing capabilities of the PS (Processing System) side to design a convolutional neural network hardware accelerator in the PL section using Verilog. By designing general-purpose modules for different types of network layers, these modules can be reused during network construction, effectively saving FPGA logic and storage resources, and facilitating task scheduling and parameter update control on the PS side.

The hardware accelerator design mainly consists of data transmission modules, data caching modules, and network layer modules. These modules are integrated and packaged into an IP core with an AXI4 bus interface for external data interaction. The overall architecture is shown in Figure 4.

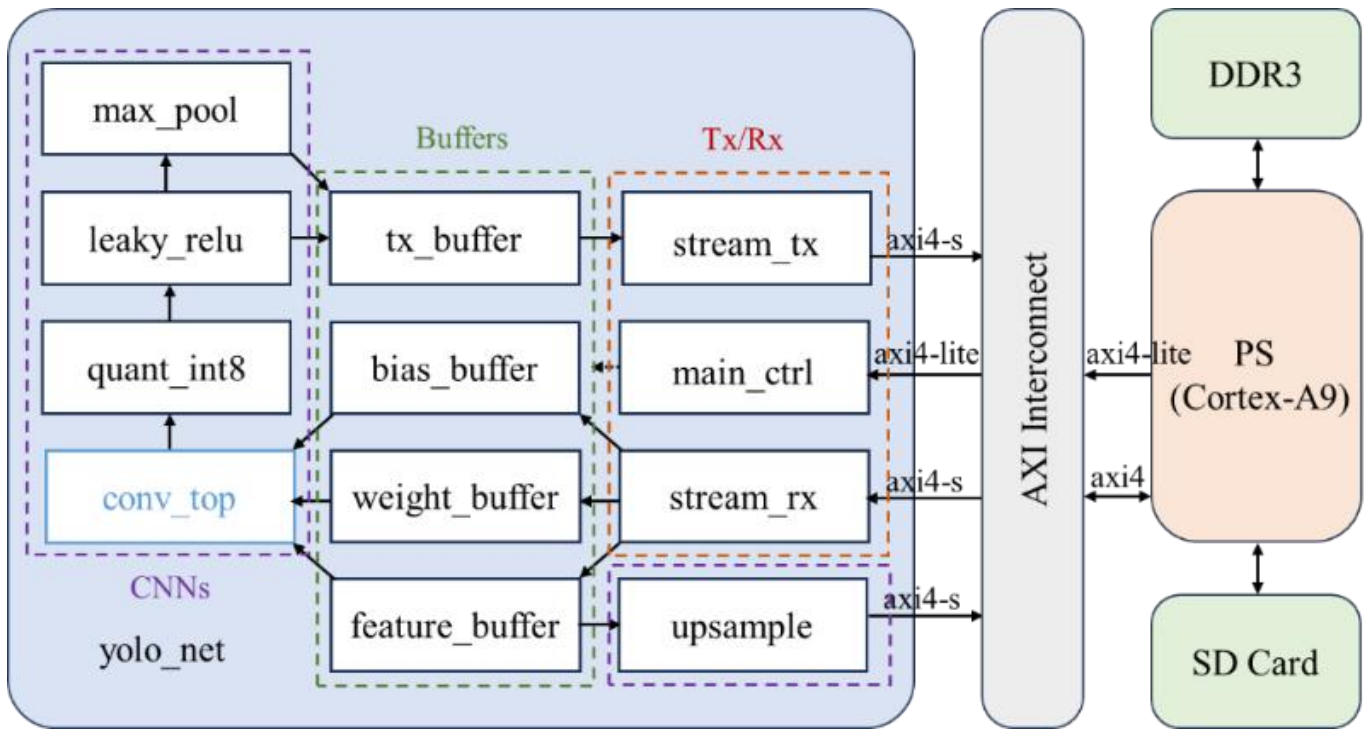

Figure 4: CNN Hardware Accelerator Block Diagram.

### 4.2 Weight Cache Optimization

Weight cache optimization is a critical component in FPGA hardware accelerator design, aimed at addressing the storage and access efficiency of convolutional layer weight data. In the YOLOv3-Tiny network, the weight data of convolution layers is accessed frequently, and directly reading from off-chip storage can increase access latency and consume significant bandwidth. To improve convolution computation efficiency, this paper designs an on-chip weight cache module that optimizes the storage and loading mechanism for the weights. Since the number of convolution kernels is often a multiple of 8, the cache module is designed based on 8 groups of RAM structures, with each group containing 8 RAM units to cache the weight parameters needed by the current convolution layer. The RAM storage array structure is shown in Figure 5.

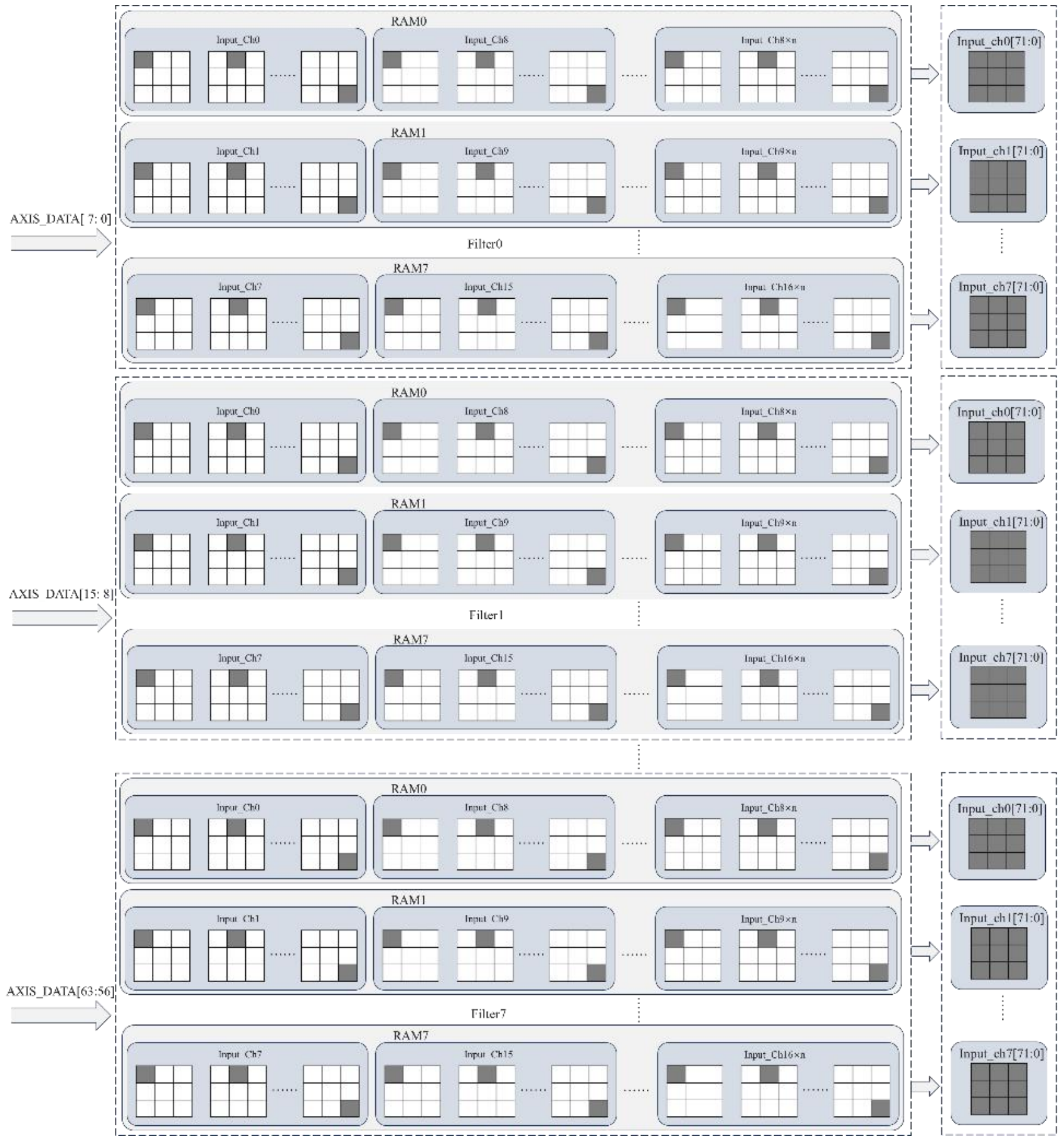

Figure 5: Weight Parameter Cache RAM Array Structure.

The 64 RAM units used by the weight parameter cache module are divided into 8 groups, with 8 RAM units in each group storing the weight parameters corresponding to a single convolution kernel. Due to the limited internal storage resources of the FPGA, each RAM has a depth of 256 and a width of 72 bits, which can store weight parameters corresponding to 256 convolution kernels and complete the data transfer for a single kernel within 9 clock cycles. The total of 64 RAM units can store 16,384 weight parameters, meeting the convolution computation requirements from Layer 0 to Layer 6. In subsequent convolution layers, as the number of input channels and convolution kernels increases, the amount of weight data also increases exponentially. When the RAM capacity is exhausted, new data will overwrite old data. A communication control module dynamically manages RAM write volume and read addresses to ensure accurate data transfer.

### 4.3 Multi-Channel Parallel Convolution

In the convolution layers of the YOLOv3-Tiny network, multi-channel parallel convolution is key to improving the computational efficiency of the hardware accelerator. This paper designs a convolution computation architecture that efficiently handles multi-channel input by utilizing DSP array reuse and pipelining mechanisms. By combining low-bit quantization and weight-sharing strategies, this architecture significantly reduces computational resource consumption while maintaining inference accuracy. Its structure is shown in Figure 6.

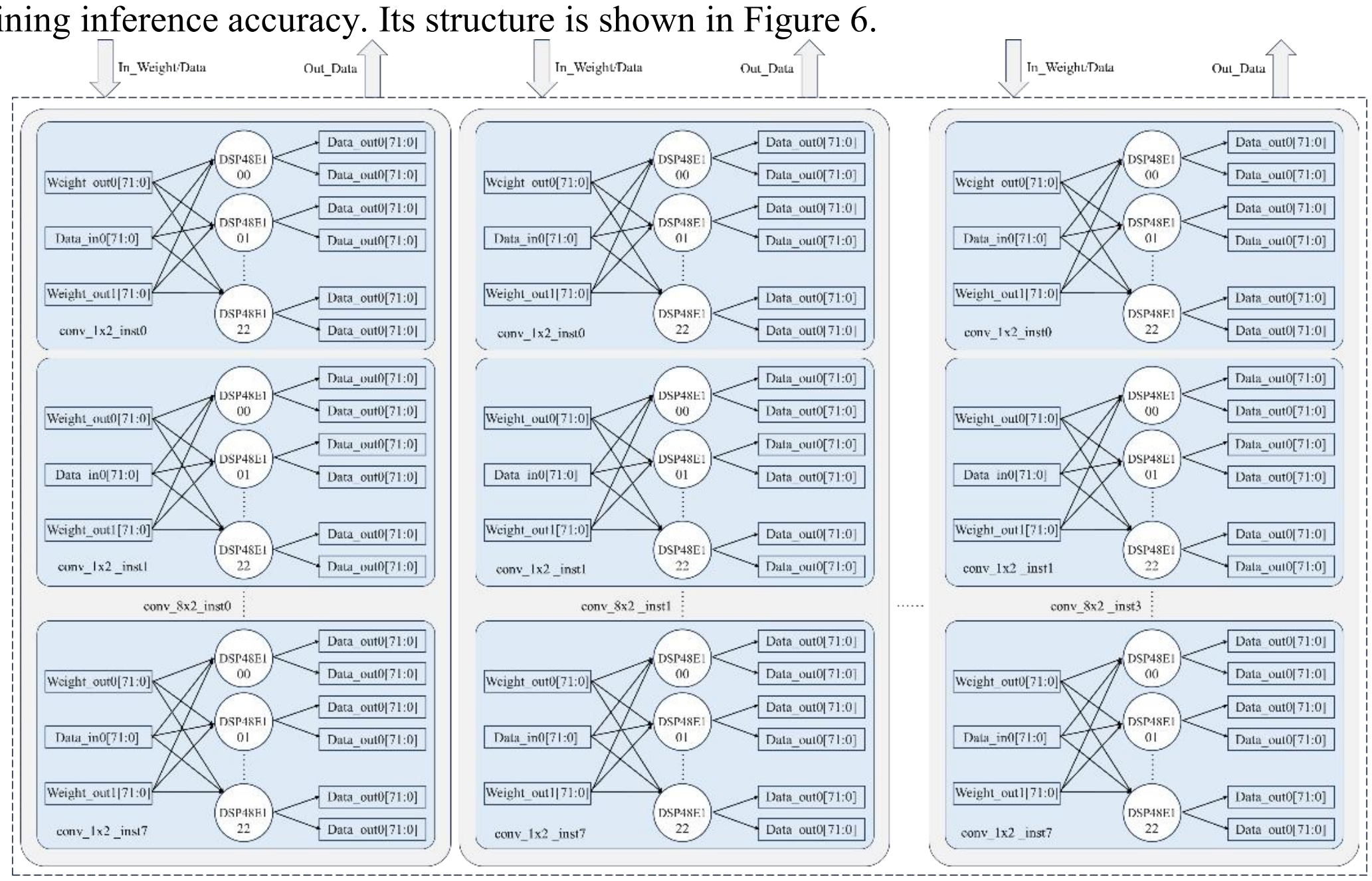

Figure 6: Convolution Computation Module DSP Array Structure.

The convolution computation module achieves multi-channel convolution operations through parallel reuse. After completing the convolution computation for 8 input channels and 8 convolution kernels in each computation batch, the input data is updated until the convolution calculation for all input channel feature maps with the 8 convolution kernels is finished. In the final batch, the bias parameter is added to complete the fusion operation of the convolution layer and BN layer on the PL side. The above computation process is repeated to perform convolution computations for multiple input channels and multiple convolution kernel weight parameters.

The DSP48E1 core in the ZYNQ-XC7Z035 is the core computational resource on the PL side. It consists of 25 18-bit multipliers, pre-adders, and multifunctional arithmetic logic units. By configuring the DSP48E1's internal functionality to $P = (A + D) \times B$ , it can process 25-bit input data $A$ and $D$ , and 18-bit input data $B$ , producing two multiplication results, $A \times B$ and $D \times B$ , in the result $P$ . By splitting the product results $P$ , the DSP dual multiplication operation can be achieved within the same clock cycle, saving 50% of DSP resources. The detailed process is shown in Figure 7.

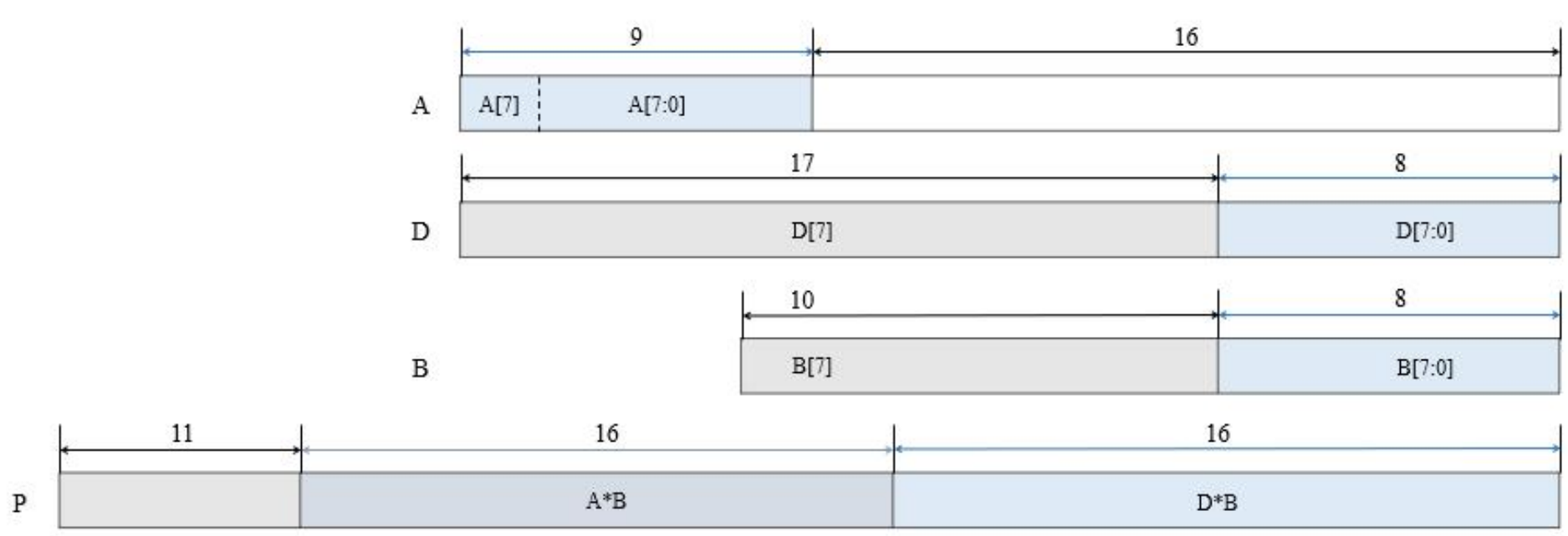

Figure 7: DSP Reuse Data Mapping Structure.

## 5 SYSTEM TESTING AND ANALYSIS

This paper trains the YOLOv3-Tiny network using the WIDER FACE: A Face Detection Benchmark dataset. This dataset contains a single object category, "Face." The trained network weights, biases, and other parameters are quantized to 8-bit fixed-point values, and the quantized results are stored on an SD card for subsequent hardware deployment.

The object detection system is deployed on the MLK_H3_7035 development board, which is equipped with the ZYNQ-XC7Z035 chip. The test results are shown in Figure 8. In the figure, the red rectangular boxes indicate the detected faces, with high prediction box coverage, demonstrating strong detection capability. The results show that the system can meet the real-time requirements of embedded devices, making it suitable for intelligent surveillance and edge computing applications in low-power environments.

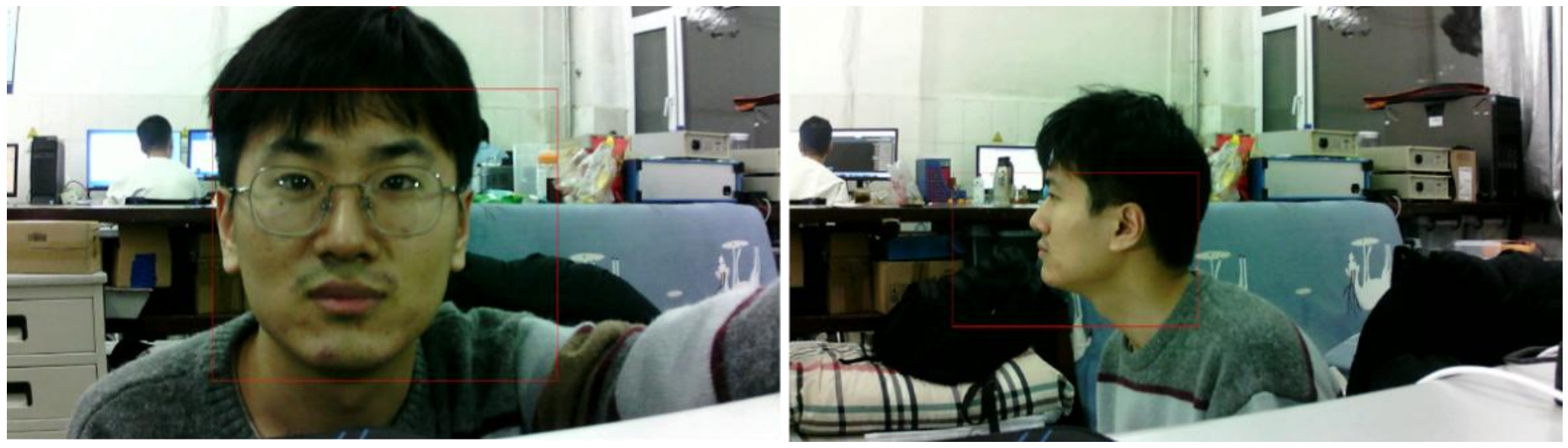

(a) Front view large object detection (b) Side view medium object detection

Figure 8: Object detection system test results.

To verify the advantages of FPGA-based embedded object detection system deployment, forward computation of the YOLOv3-Tiny network was performed on both CPU and GPU platforms. The test devices included an Intel i5-11400F (CPU) and a GeForce RTX3060 (GPU). The experimental results from different hardware platforms are compared in Table 1. The forward inference latency for the YOLOv3-Tiny network on the ZYNQ-XC7Z035 platform was 0.211 seconds, outperforming the CPU and most embedded platforms, although slightly slower than the GPU. Compared to platforms such as Raspberry Pi and ARM, the ZYNQ-XC7Z035 platform shows significant performance advantages. In addition, the design based on the ZYNQ-XC7Z035 platform achieved a throughput of 26.41 GOPS, with a power consumption of 2.613W and an power efficiency of 10.11 GOPS/W. This is higher than the CPU, GPU, and most embedded platforms, and only slightly lower than highly optimized ASIC platforms, offering high computational performance while maintaining low power consumption. From this, it is evident that the ZYNQ-XC7Z035 platform, using FPGA as the main computational structure, offers high energy efficiency, strong parallel computing capability, and high customizability, making it well-suited for embedded object detection system development and application.

Table 1: Comparison of Lightweight Convolutional Neural Network Hardware Acceleration Performance Across Platforms.

| Design | [9] | [10] | [11] | This Work | This Work | This Work |
|---|---|---|---|---|---|---|
| Platform | ARM Cortex-A9 | ASIC | Raspberry Pi 4B | Intel i5-11400F | GeForce RTX3060 | ZYNQ-7035 |
| Model | YOLOv3-Tiny | YOLOv5-s | YOLOv5-Lite | YOLOv3-Tiny | YOLOv3-Tiny | YOLOv3-Tiny |
| Frequency(Hz) | 667M | 100M | 1.5G | 4.4G | 1.3G | 200M |
| Precision(bit) | 16 | 16 | 32 | 32 | 32 | 8 |
| Latency(s) | 142.857 | 0.245 | 0.250 | 0.698 | 0.057 | 0.211 |
| Throughput(Gops) | 0.04 | 65.2 | 3.6 | 8.023 | 98.26 | 26.41 |
| Power Efficiency (Gops/W) | 0.013 | 318.049 | - | 0.123 | 0.578 | 10.11 |

To further validate the performance of the hardware accelerator designed in this paper, it is compared with other FPGA implementation solutions. The specific performance is shown in Table 2. On the FPGA platform, the YOLOv3-Tiny network deployed on the ZYNQ-XC7Z035 platform can still achieve good timing convergence with a high clock frequency of 200 MHz. By applying 8-bit low-bit quantization, the network's parameter size is significantly compressed, achieving faster inference speed and an higher power efficiency with minimal FPGA resource usage, effectively meeting the expected research goals.Compared to other solutions, the proposed design achieved an inference latency of 0.211 seconds, which is 75.58% faster than similar designs, and it meets real-time requirements. The power efficiency reached 10.11 GOPS/W, which is at least 29.45% better than similar designs, demonstrating low power consumption and high performance. The hardware resource consumption is lower, with a reduction of up to 51.94% compared to similar designs, significantly saving hardware resources while ensuring performance. This result shows that the design achieves a good balance between high performance and low power consumption.

Table 2: Performance Comparison of YOLOv3-Tiny Hardware Acceleration Schemes Based on FPGA.

| Design | [9] | [12] | [13] | [14] | This Work |
|---|---|---|---|---|---|
| Platform | ZYNQ-7020 | ZYNQ-7020 | ZYNQ-7035 | ZYNQ-7035 | ZYNQ-7035 |
| Frequency(Hz | 100M | 100M | 100M | 100M | 200M |

| Design ) | [9] | [12] | [13] | [14] | This Work |
|---|---|---|---|---|---|
| Precision(bit) | 16 | 16 | 16 | 16 | 8 |
| DSP | 174 | 160 | 366 | 485 | 304 |
| LUT | 28.8k | 25.9k | 63.6k | 61.7k | 20.1k |
| FF | 49.3k | 46.7k | 32.5k | - | 24.8k |
| Latency(s) | 0.611 | 0.532 | 0.864 | 0.192 | 0.211 |
| Throughput(Gops) | 9.12 | 10.45 | 13.26 | 28.99 | 26.41 |
| Power Efficiency (Gops/W) | 4.26 | 3.11 | 4.01 | 7.81 | 10.11 |

## 6 CONCLUSION

This paper addresses the efficiency and low power consumption requirements for embedded object detection by designing and implementing an FPGA-based YOLOv3-Tiny hardware acceleration system. In this study, the design incorporates techniques such as low-bit quantization, batch normalization fusion, and table lookup mapping optimization. By combining lightweight network optimization with hardware acceleration, an efficient hardware acceleration framework is designed. The data flow scheduling, storage resource allocation, and computational unit utilization are optimized, achieving a forward inference latency of 0.211 seconds and an power efficiency of 10.11 GOPS/W on the ZYNQ-XC7Z035 platform, which is at least 29.45% better than similar designs. Experimental results show that this solution excels in inference speed, hardware resource usage, and power consumption control. It meets the real-time and low-power requirements of embedded scenarios, providing an efficient solution for object detection tasks in resource-constrained environments. Future research could further explore lightweight convolutional neural network hardware acceleration schemes based on modular designs to support more network architectures, while balancing custom design and development efficiency, to promote the widespread application of FPGA hardware acceleration technology in embedded object detection systems.

## ACKNOWLEDGMENTS

We gratefully acknowledge the financial support from the Science and Technology Development Plan Project of Jilin Province (Grant No. 20240302015GX).